# Continued Monitoring of the Conditioning of the Fermilab Linac 805 MHz Cavities*

E. McCrory, T. Kroc, A. Moretti, M. Popovic, Fermilab, Batavia, IL 60510, USA

*Abstract*

We have reported previously on the conditioning of the high-gradient accelerating cavities in the Fermilab Linac [1, 2, 3]. Automated measurements of the sparking rate have been recorded since 1994 and are reported here. The sparking rate has declined since the beginning, but there are indications that this rate may have leveled off now. The X-rays emitted by the cavities are continuing to decrease.

## 1. INTRODUCTION

Fermilab commissioned the seven, high-gradient 805 MHz RF accelerating modules in 1993. In order to achieve the desired acceleration, gradients of up to 8 MV/m were required, which led to maximum surface gradients of nearly 40 MV/m. These high fields caused some concern about RF breakdown leading to beam loss and to excessive X-ray exposure.

After seven years, it seems that the change in the rate of these breakdowns has stabilized at a level well below the original specifications: a lost beam rate due to RF breakdown/sparking of 0.1% or less.

## 2. OVERVIEW OF MEASUREMENTS

Automated measurements of the sparking rate of each of the seven 805 MHz RF cavities in the 400 MeV Fermilab Linac have been collected since April 1, 1994. Also, we have automatically recorded the number of beam pulses lost each day, presumably due to RF breakdown in one or more of the cavities, beginning in January 1994.

We have measured the X-ray production rate as a function of the power levels in one cavity on several occasions over these years.

### 2.1. Sparking Rate

The sparking rate has been measured continually at the 15 Hz repetition rate of our RF system. These data have been recorded daily. We have accumulated 1893 days of data (82% of the available days). We record the number of RF pulses for each of the seven 805 MHz cavities and the number of times an RF pulse at that cavity was ruined by an RF breakdown/spark. We have experimented with various ways of detecting sparks in the cavities, and have determined that watching for abnormal reverse power from the cavity is the most reliable. We tried for approach pressure was relatively high and stable. But now, better vacuum conditions, coupled with regular, small vacuum bursts unrelated to spark activity make the spark-induced vacuum activity harder to identify. The ratio of these two methods of counting varies by about a factor of two from day to day, with an average ratio of 2 reverse-power-only count for every reverse-power-and-vacuum-activity count. The data we present here are for the reverse-power-only method.

#### 2.1.1. The Overall Rate

Table 1 shows the median number of sparks per day for each of the years we have been accumulating data. Most days have about $1.296 \times 10^6$ RF pulses per cavity.

| Year | Days | M1  | M2 | M3  | M4 | M5 | M6 | M7 |
|------|------|-----|----|-----|----|----|----|----|
| 1994 | 262  | 85  | 5  | 112 | 35 | 42 | 19 | 4  |
| 1995 | 324  | 57  | 8  | 66  | 13 | 26 | 13 | 4  |
| 1996 | 318  | 124 | 40 | 44  | 14 | 27 | 14 | 3  |
| 1997 | 289  | 92  | 24 | 26  | 7  | 19 | 6  | 1  |
| 1998 | 196  | 68  | 8  | 7   | 6  | 8  | 6  | 1  |
| 1999 | 295  | 11  | 10 | 17  | 6  | 7  | 6  | 2  |
| 2000 | 141  | 29  | 18 | 19  | 5  | 7  | 7  | 1  |

**Table 1.** The median number of sparks per day.

The "Days" column represents the number of days counted, based on the total number of RF pulses recorded that year. Note that 1998 had only 196 equivalent days—this is due to a series of major shutdowns in the Linac that year. The jumps in the numbers in this Table, particularly between 1995 and 1996 in Modules 1 and 2, correspond to increasing the length of time the RF is at full value (the "pulse length").

There is no indication that sparking is correlated among the cavities. So, one would expect that the sum of the values in each row would represent the median number of sparks in the entire Linac per day.

We currently expect about 86 ± 32 sparks in the Linac per day. This is the median number of sparks, summed over all cavities, ignoring possible correlations. The error bars represent the quadrature sum of the standard deviation on the number of sparks per cavity, per day. This is a rate of $(6.6 \pm 2.5) \times 10^{-5}$ sparks per RF cycle, or about one spark every 17 minutes of operation. This is well below the original specification of 1 spark in the Linac for every 1000 RF cycles.

#### 2.1.2. Rates Per Cavity

The sparking rate of a cavity depends on many things, and cold, startup effects often dominate getting a clean reading of the rate per day. (The RF systems are interlocked, so coming out of an enclosure access often causes small problems, which are generally manifested by high reverse power that are not necessarily associated with sparking.) Module 3 seems to have been the most stable over these years, so we present the sparking rate per pulse per day in Figure 1 for Module 3. The other modules show similar characteristics, but because we have done more experimentation with the pulse length on them, the data are not as clear.

---

* Work supported by the US Department of Energy, contract # DE-AC02-76CH0-3000.

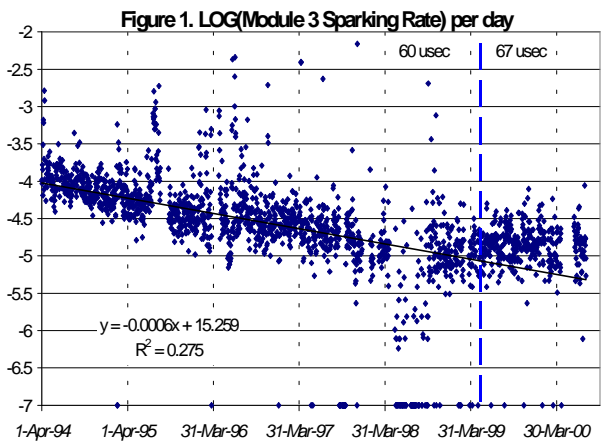

Figure 1. LOG(Module 3 Sparking Rate) per day

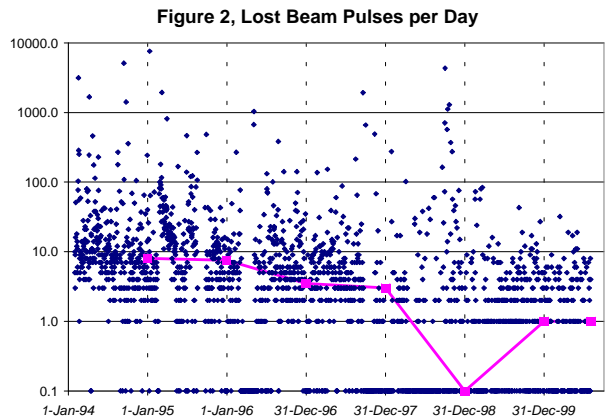

Figure 2, Lost Beam Pulses per Day

The most striking feature of this graph is that the sparking rate has steadily declined for the entire measurement period, and is only now beginning to show signs of leveling off. (Note that one spark per day would be a sparking rate of just below $1 \times 10^{-6}$, or "-6" on this graph).

The fit to these data for Module 3 says that in 1700 days, the sparking rate has decreased by a factor of ten.

We have experimented with changing the RF pulse length on many of the cavities. We changed the pulse length on Module 3 in June of 1999 from 60 to 67 microseconds. According to our previous paper [2], we would expect the sparking rate to increase by a factor proportional to the fourth power in the pulse length. $(67/60)^4 = 1.55$, which is consistent with the data presented here.

Prior to the pulse length change, it appears that the sparking rate on Module 3 may have begun to level off at a rate of one spark every $10^5$ RF pulses. The other modules have a similar behavior, although it is difficult to factor out the effect of the lengthening of the RF pulse. We will continue to monitor the sparking rate and report again in a few years.

### 2.2. Lost Beam

We also began counting the number of lost beam pulses per day in 1994. The algorithm for determining this, while not ideal, is reasonable: At the repetition rate of the RF systems (15 Hz), we look for a beam pulse by watching the current on the beam toroid at the beginning of the 805 MHz section (at 116 MeV). If the beam current is above 20 mA, then this cycle is a beam pulse. If, then, the beam current out of the end of the linac (400 MeV) is less than 20 mA, we record this as a lost beam pulse. This records all sparks that result in a loss of beam, but it also captures the occasional beam pulse during routine tuning where the input current is just over 20 mA and the output current is just under that level. We estimate that on days with ten or more lost beam pulses, one can reasonably expect that one or two are from this effect.

The data for the number of lost beam pulses are shown in Figure 2. The line represents the median number of lost beam pulses per day for the year, calculated on the last day of the year. The number of lost beam pulses per day was significantly larger in 1994 than it is now (an average of 64.4 and a median of 9 with a standard deviation of 398 in 1994 versus $2.1 \pm 3.8$ (median = 1) now). Zero is represented as 0.1 on this log graph. The median number of lost pulses per day in 1998 was zero because we were down for a large fraction of that year.

With 30000 beam pulses per day, we would expect $1.98 \pm 0.74$ lost beam pulses per day due to the RF breakdown rate of $6.6 \times 10^{-5}$. Since we measure between 1 and 2 lost beam pulses per day, we can conclude that the presence of beam does not have an appreciable effect on the sparking rate in our cavities. In [1], we reported that there is a 20% increase in the sparking rate during beam. The statistics do not justify this conclusion now.

### 2.3. X-Ray Measurements

We have measured the X-ray levels at each of the four sections of Module 5 on several occasions: once when it was first commissioned, once for the 1996 paper, and once again now. The data are shown in Figure 3.

The 1992 data were taken with a single detector placed approximately four feet transversely from the center of the module, between sections 2 and 3. The rest of the data were taken with four detectors placed approximately 1 foot transversely from the center of each of the four sections of the module. The 1992 data have been multiplied by four (assuming a quasi-line source) to suggest the proper relationship to the other data that have not been transformed.

We fit the data from each detector to the Fowler-Nordheim equation for an RF field that describes enhanced field emission [4].

$$j_F = \frac{5.7 \times 10^{-12} \times 10^{4.52\phi^{-0.5}}}{\phi^{1.75}} A_e (\beta E_0)^{2.5} \exp\left(-\frac{6.53 \times 10^9 \times \phi^{1.5}}{\beta E_0}\right)$$

where $\phi$ is the work function in eV, $E_0$ is the macroscopic surface field in V/m, $A_e$ is the area of the emitting site, and $\beta$ is the enhancement factor. The enhanced field emission is presumed to occur due to some mechanism with area $A_e$ that magnifies the local electric field by a factor of $\beta$. We fit our data to this form using MINUIT [5] with the free parameters being $\beta$ and a term proportional to $A_e$. These data are shown in table 2.

| Data set | Area term | $\beta$ | $\chi^2$/DoF |
|---|---|---|---|
| 1992 | $1.0 \times 10^{-12} \pm 4 \times 10^{-14}$ | 274±1 | 433.1 |
| 1996_1 | $8.8 \times 10^{-9} \pm 3 \times 10^{-10}$ | 128±0.3 | 12.8 |
| 1996_2 | $3.6 \times 10^{-9} \pm 1 \times 10^{-10}$ | 140±0.3 | 52.4 |
| 1996_3 | $2.2 \times 10^{-9} \pm 8 \times 10^{-10}$ | 133±2.9 | 0.23 |
| 1996_4 | $1.7 \times 10^{-8} \pm 3 \times 10^{-10}$ | 129±0.2 | 496.9 |
| 2000_1 | 0.32±0.009 | 59.7±0.05 | 7.7 |
| 2000_2 | 0.54±0.004 | 59.6±0.02 | 43.8 |
| 2000_3 | 0.23±0.013 | 59.6±0.1 | 2.8 |
| 2000_4 | 0.74±0.003 | 59.9±0.01 | 336.2 |

In these data we see a reduction of beta of a factor of two every four years. On the other hand, the area term has increased by many orders of magnitude. One can surmise that high beta sites are being removed and thereby eliminating their domination over lower-beta but larger-area sites. Notice, however, the behavior of the 1996 and 2000 data above 37 - 38 MV/m where the more recent data shows higher readings than the previous data even though the detector geometries were the same. This would imply that the emitting area increased over time.

## CONCLUSION

The sparking rate in the Fermilab 805 MHz, 400 MeV Linac has reduced to approximately one cavity spark for every 17 minutes of RF operation. There are indications that it may continue to decrease even further. The lost beam rate is approximately equal to the sparking rate, indicating that the presence of our beam has no impact on the sparking rate.

Analyzing the X-ray data with the theory of enhanced field emission shows continued reduction of the enhancement factor. However, the X-ray levels at the operating gradient have remained the same or even increased slightly. This implies an additional change, possibly an increase in the emitting area.

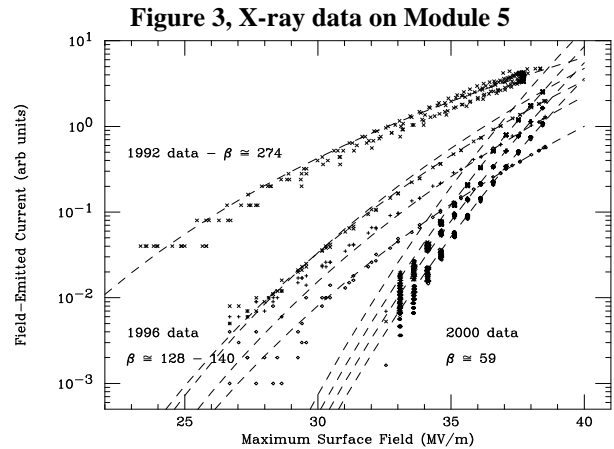

Figure 3, X-ray data on Module 5